\documentclass[twocolumn,american,prl]{revtex4}
\usepackage[latin1]{inputenc}
\usepackage{graphicx}

\makeatletter

\providecommand{\tabularnewline}{\\}

\usepackage{ae,aecompl}

\usepackage{babel}
\makeatother
\begin{document}

\title{Soliton topology versus discrete symmetry in optical lattices}

\author{Yaroslav V. Kartashov,$^{1}$ Albert Ferrando,$^{2}$ Alexey A. Egorov,$^{3}$
and Lluis Torner$^{1}$}

\affiliation{$^{1}$ICFO-Institut de Ciencies Fotoniques,
Universitat Politecnica de Catalunya, Barcelona, Spain.}

\affiliation{$^{2}$Interdisciplinary Modeling Group, Departament
d'\`{O}ptica, Universitat de Val\`{e}ncia. Dr. Moliner, 50.
E-46100 Burjassot (Val\`{e}ncia), Spain.}

\affiliation{$^{3}$Physics Department, M. V. Lomonosov Moscow
State University, 119899, Moscow, Russia.}

\date{\today{}}

\begin{abstract}
We address the existence of vortex solitons supported by
azimuthally modulated lattices and reveal how the global lattice
discrete symmetry has fundamental implications on the possible
topological charges of solitons. We set a general ``charge
rule'' using group-theory techniques, which holds for all lattices
belonging to a given symmetry group. Focusing in the case of
Bessel lattices allows us to derive also a overall stability rule
for the allowed vortex solitons.
\end{abstract}


\maketitle

Vortex solitons in nonlinear systems (for
a review see \cite{ProgOptics}) characterized by a discrete symmetry have been
numerically predicted in two-dimensional arrays of
evanescently coupled waveguides \cite{malomed-pre64_026601},
harmonic refractive index gratings imprinted in cubic media
\cite{yang_and_baizakov}, as well as in photonic
crystal fibers with defects \cite{ferrando-oe12_817}. Because of
the imprinted refractive-index modulation, such vortices can be
made stable in contrast to their ring-shaped counterparts in
uniform focusing media. Recently, vortices having unit topological
charge have been experimentally observed in optically-induced
lattices in photorefractive media
\cite{neshev_and_fleischer}. The very
refractive index modulation causing the stabilization of vortex
solitons simultaneously imposes restriction on the possible
topological charges of the vortices dictated by the finite order
of allowed discrete rotations \cite{ferrando}. A
corollary of such result is that the maximum
charge of stable symmetric vortex in two-dimensional square lattices is one.

However, square lattices are just one particular example of
guiding structures accessible for experimental exploration.
Another interesting class of such structures with a new global
rotational symmetry is constituted by azimuthally modulated
lattices, also offering a wealth of new opportunities. For example, in such
lattices the order of rotational symmetry may be higher than 4, in
contrast to square lattices, a property that has direct
implications in the possible topological charges of symmetric vortex
solitons supported by such lattices.

In this Letter we explore the connection existing between the
lattice discrete symmetry and the topology of the allowed vortex
solitons by means of a general group-theory approach. We find
that azimuthally modulated lattices imprinted in focusing
medium can support symmetric vortex solitons carrying phase dislocations
with topological indices higher that one. The higher
the symmetry order of the lattice, the higher the allowed vortex
topological charge. We also address the
stability of the allowed vortex soliton families, taking as a particular
example the case of azimuthally modulated optically-induced Bessel lattices.

Optical lattice induction in anisotropic nonlinear materials
introduced in \cite{fleischer_and_neshev} opens broad prospects
for creation of reconfigurable refractive index landscapes with
different types of nondiffracting beams, including Bessel beams
\cite{kartashov_et_al}. Accurate approximations of Bessel beams
can be generated experimentally in a number of ways. Known
techniques include illumination of annular slit in the focal plane
of a lens, conical axicons, as well as more complicated
interferometric and holographic techniques (see
\cite{arlt_and_dholakia} and references therein). Thus,
azimuthally modulated lattices of any desired order can  be
optically induced by higher-order Bessel beams
\cite{kartashov_et_al}. Because of the diffractionless nature of
Bessel beams, they are to be launched collinearly along the
anisotropic nonlinear material (e.g., photorefractive crystal)
with a polarization orthogonal to the soliton beams, to make use
of incoherent vectorial interactions \cite{fleischer_and_neshev}.
The concept can be extended to all relevant physical settings,
including Bose-Einstein condensates.

Our starting point is the paraxial nonlinear equation for the
complex field amplitude $q$ describing the propagation of light in
azimuthally modulated lattice: \begin{equation} i\frac{\partial
q}{\partial\xi}=-\frac{1}{2}\left[\frac{\partial^{2}}{\partial\eta^{2}}+\frac{\partial^{2}}{\partial\zeta^{2}}\right]
q-q|q|^{2}-pR(\eta,\zeta)q.\label{eq:evolution_equation}\end{equation}
The longitudinal $\xi$ and transverse $\eta$, $\zeta$ coordinates
are scaled to the diffraction length and to the input beam width,
respectively. The parameter $p$ accounts for the depth of the
refractive index profile, whereas the function $R(\eta,\zeta)$
describes the lattice profile. In the particular case of optical
lattice induced by higher-order Bessel beam the refractive index
profile features beam's intensity distribution
$R_{n}(\eta,\zeta)=J_{n}^{2}[(2b_{\mathrm{lin}})^{1/2}r]\cos^{2}(n\phi)$,
where $r=(\eta^{2}+\zeta^{2})^{1/2}$, $\phi$ is the azimuth angle,
and $b_{\mathrm{lin}}$ is the parameter that sets the transverse
lattice scale. In the particular case of optical lattice induction
in SBN crystal biased with dc electric field $\sim 10^{5}$ V/m,
for laser beams with width 10 $\mu$m the propagation distance
$\xi\sim 1$ corresponds to 1 mm of actual crystal length, while
amplitude $q\sim 1$ corresponds to peak intensity about $50 mW/cm^{2}$.
The intensity of lattice-creating wave is of the same
order and can be varied to tune $p$ in a broad range.

The realization of rotational symmetry in azimuthally modulated
lattices is different depending on the order of modulation $n$.
Lattices of lowest-order with $n=0$ show perfect continuous
rotational symmetry since the function
$R_{0}(\eta,\zeta)=R_{0}(r)$ has no dependence on $\phi$ \cite{kartashov_et_al}. In terms of group
theory, the symmetry group of a lattice of zero order is $O(2)$.
The azimuthal modulation $R_{n}(r,\phi)\sim\cos^{2}(n\phi)$
changes the symmetry group associated to the rotation
transformations. The rotational symmetry group of a lattice of
order $n$ is given by the discrete point-symmetry group
$\mathcal{C}_{2n,v}$, corresponding to discrete rotations of angle
$\epsilon_{n}=\pi/n$ with respect to a rotation axis perpendicular
to the plane and intersecting it at the origin ($R_{n}(r,\phi+\epsilon_{n})=R_{n}(r,\phi)$)
as well as to specular reflections with respect to a number of planes containing
the rotation axis \cite{hamermesh64}.
This fact has strong implications in the form of possible
symmetric vortex solutions of Eq.(\ref{eq:evolution_equation})
$q(\eta,\zeta,\xi)=\left[u(\eta,\zeta)+iv(\eta,\zeta)\right]\exp(ib\xi)$,
where $u$ and $v$ are real and imaginary parts, respectively,
$b$ is a propagation constant, and $m$ describes topological winding
number of complex field $q$ that can be defined by the circulation of
the gradient of the field phase $arctan(v/u)$ around the singularity at $\eta,\zeta=0$.

Individual vortices are characterized by a phase singularity,
accompanied by a single point of zero amplitude. Discrete
rotations are defined with respect to this point, so that all our
group theory arguments will apply to individual vortices and not
to bound states or clusters of individual vortices owning
intrincated phases featuring edges and multiple phase
singularities. In this context, individual vortex solutions appear
as doubly-degenerated pairs belonging to the two-dimensional
representations of $\mathcal{C}_{2n,v}$ (or of its subgroups) and
they are characterized by the index representation $m$. The charge
of these solutions is exactly provided by the index representation
$m$ \cite{ferrando}. For every value of $m$ one finds a
degenerated vortex-antivortex pair, with charges $m$ and $-m$,
respectively. Since $m$ is limited by symmetry constraints, an
upper bound for the values of permitted vortex charges is
established if the system enjoys a rotational symmetry of finite
order $N$ \cite{ferrando}: $0<m<N/2$ (for even $N$). Since the
symmetry group of a lattice of order $n$ is $\mathcal{C}_{2n,v}$,
its symmetry order is $N=2n$ and thus even. Consequently, we
obtain one of the central results of this Letter in the form of a
{}``charge rule'' for the allowed values of vortex charges in a
lattice of order $n$:\begin{equation} 0<m\leq
n-1\label{eq:charge_rule}\end{equation}

We stress that this rule is obtained on the basis of general
symmetry arguments and is applicable for a whole class of
azimuthally modulated lattices, irrespectively of the details of
their local shape and thus method of realization. Some
implications of \ref{eq:charge_rule} are readily apparent. For
example, azimuthally modulated lattices of high orders allow
generation of symmetric vortex solitons with topological charges
higher than one, in contrast to square lattices.

\begin{figure}
\includegraphics[%
  scale=0.50]{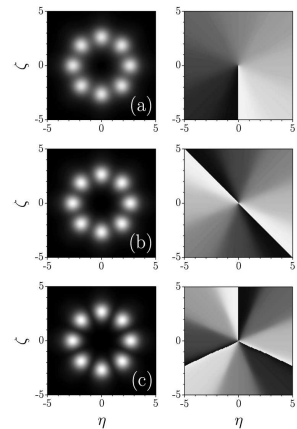}

\caption{Amplitude (left column) and phase (right column) distributions for
the vortex solitons with charges 1 (a), 2 (b), and 3 (c), supported
by the fourth-order Bessel lattice at $p=28$. All solitons correspond
to $b=1.6$. In left column bright regions correspond to high light
intensities, dark regions correspond to low intensities.\label{cap:Amplitude_and_phase_vortices}}
\end{figure}

We performed a comprehensive numerical analysis of vortex soliton
solutions of Eq.(\ref{eq:evolution_equation}) for the case of
optically induced Bessel lattices to confirm the ``charge rule''.
We searched for symmetric stationary vortex profiles with a
relaxation method using as initial guess functions with a phase
profile $exp(im\phi)$ carrying phase-singularity with charge $m$.
The numerical results confirmed the ``charge rule'' in all cases.
A summary of this analysis is presented in Table \ref{cap:Table}.
No higher-order vortex solitons have been found to exist above the
maximum limit given by Eq. (\ref{eq:charge_rule}). The amplitude
and phase distributions for allowed vortices with charges 1, 2 and
3 in a Bessel lattice of fourth order are shown in Fig.
\ref{cap:Amplitude_and_phase_vortices}. The presence of focusing
nonlinearity tends to increment the localization effect generating
the typical patterns of bright well-localized spots. Notice, that
in sufficiently deep lattices at fixed $b$ and $p$ the bright
spots in the vortex intensity distribution become more pronounced
with increase of vortex charge $m$, while radius of the vortex
remains almost unchanged since it is mainly determined by
transverse lattice scale. This is in contrast to vortices in
uniform medium that broaden notably with increase of $m$ at fixed
$b$.

\begin{figure}
\includegraphics[%
  scale=0.65]{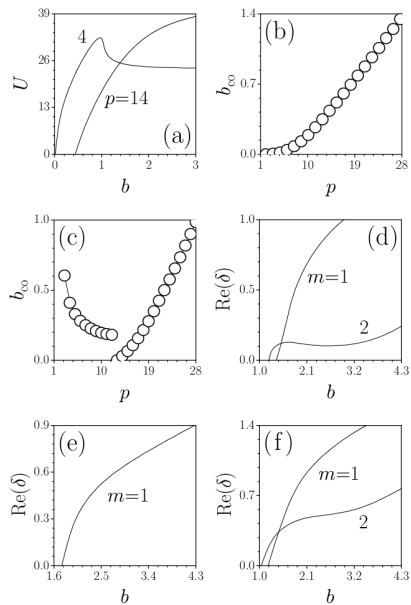}

\caption{(a) Energy flow $U$ versus propagation constant $b$ for a vortex
soliton with $m=1$. Propagation constant cutoff $b_{\mathrm{co}}$
versus lattice depth $p$ for vortex solitons with $m=1$ (b) and
$m=3$ (c). (d) Real part of perturbation growth rate $\delta$ versus
propagation constant at $p=28$. Panels (a)-(d) correspond to fourth-order
Bessel lattice. Real part of perturbation growth rate versus propagation
constant for vortex solitons in (e) third- and (f) fifth-order lattices
at $p=28$.\label{cap:diagrams_vortex_charge_1}}
\end{figure}

The properties of vortex solitons supported by fourth-order
azimuthally modulated Bessel lattice are summarized in Figs.
\ref{cap:diagrams_vortex_charge_1}(a)-(d). The vortex soliton is
characterized by its energy flow $U=\int\int|q|^{2}d\eta d\zeta$. The
dependence $U(b)$ for a lowest-order vortex soliton with
$m=1$ is shown in Fig. \ref{cap:diagrams_vortex_charge_1}(a). The
different behavior of the $U(b)$ curve for a shallow ($p=4$) or
deep ($p=14$) lattice indicates the importance of the confining
properties of the Bessel lattice. The energy flow is a
non-monotonic function of the propagation constant for small
lattice depths ($p=4$) whereas for large values of $p$
(approximately more or equal than 6) it monotonically increases
with $b$. The deeper is the lattice the more pronounced is the
azimuthal modulation of the vortex intensity profile.

In contrast, $m=1$ vortices in shallow lattices may undertake an abrupt change of
behavior at a critical value of the propagation constant (see the
inflection point of the $U(b)$ curve at $p=4$). With increase of
energy flow , bright spots forming such vortices may eventually merge
into a single weakly modulated bright ring, since the confining
effect of the shallow Bessel lattice is not sufficient to ensure
the trapping of light in the higher refractive index regions.
At a given value of $p$ the energy flow of the vortex soliton
vanishes at certain cutoff $b_{\mathrm{co}}$ on propagation constant (Fig.
\ref{cap:diagrams_vortex_charge_1}(a)). Close to the cutoff, vortex with charge $m=1$
broadens drastically in shallow lattices, while in deep lattices
brigth well-localized spots are always resolvable in vortex
intensity profile. As shown in Fig.
\ref{cap:diagrams_vortex_charge_1}(b), for vortex with charge
$m=1$ the cutoff is a monotonically increasing function of the
lattice depth $p$. For higher-order vortices with charges $m>1$
the dependence $b_{\mathrm{co}}(p)$ may be discontinuous (Fig.
\ref{cap:diagrams_vortex_charge_1}(c)). Thus, in shallow lattices
higher-order vortices cease to exist in the cut-off without any
topological transformation (left branch of the
$b_{\mathrm{co}}(p)$ curve), while in deep enough lattices they
drastically broaden in the cutoff where vortex energy flow
vanishes (right branch).

Next we conducted a detailed linear stability analysis for the
allowed vortex soliton families. We searched for perturbed solutions  of
Eq.(\ref{eq:evolution_equation}) in the form $
q(\eta,\zeta,\xi)=\left[u(\eta,\zeta)+u_{p}(\eta,\zeta,\xi)+iv(\eta,\zeta)+iv_{p}(\eta,\zeta,\xi)\right]\exp(ib\xi)$,
where $u_{p}$ and $v_{p}$ are real and imaginary parts of perturbation,
respectively. The linearized evolution equations for perturbation components are

\[
\frac{\partial u_{p}}{\partial\xi}=-\frac{1}{2}\Delta_{\perp}v_{p}+bv_{p}-\left[2uvu_{p}+(3v^{2}+u^{2})v_{p}\right]-pRv_{p}
\]

\[
-\frac{\partial v_{p}}{\partial\xi}=-\frac{1}{2}\Delta_{\perp}u_{p}+bu_{p}-\left[2uvv_{p}+(3u^{2}+v^{2})u_{p}\right]-pRu_{p}
\] where $\Delta_{\perp}$ stands for the transverse Laplacian. We
solved this system numerically by means of a split-step Fourier
method with noisy initial conditions to get perturbation profiles
and their growth rates.

The results of the stability analysis are summarized in Figs.
\ref{cap:diagrams_vortex_charge_1}(d)-(f) and in Table
\ref{cap:Table} for lattices with $n$ up to 6. Thus, in fourth-
and fifth-order Bessel lattices (Figs.
\ref{cap:diagrams_vortex_charge_1}(d) and
\ref{cap:diagrams_vortex_charge_1}(f)), vortices with $m=1$ and 2
were found to be unstable in the entire domain of their existence,
while in third-order lattice the unstable vortex carries charge
$m=1$ (Fig. \ref{cap:diagrams_vortex_charge_1}(e)). Both
exponential and oscillatory instabilities are encountered for such
vortices. In contrast, third-order lattice can support stable
vortex with charge $m=2$, fourth-order lattice supports stable
vortex with $m=3$, and fifth-order lattice supports stable
vortices with $m=3$ and 4. In shallow lattices the domains of
existence of such vortices feature multiple stability and
instability windows, while for deep enough lattices such vortices
become stable in the entire domain of their existence. The
physical origin of such stabilization is the higher confinement of
radiation in regions with higher refractive index that reduces the
strength of nonlinear interactions of bright spots forming the
vortex. In experiment vortex stabilization would become apparent
upon gradual increase of intensity of the lattice-creating beam.

\begin{table}
\begin{tabular}{|c|c|c|c|c|c|}
\hline
Lattice order&
\multicolumn{5}{c|}{Available charges and stability status}\tabularnewline
\hline
\hline
$n=2$&
$m=1$&
&
&
&
\tabularnewline
&
\textbf{stable}&
&
&
&
\tabularnewline
\hline
$n=3$&
$m=1$&
$m=2$&
&
&
\tabularnewline
&
unstable&
\textbf{stable}&
&
&
\tabularnewline
\hline
$n=4$&
$m=1$&
$m=2$&
$m=3$&
&
\tabularnewline
&
unstable&
unstable&
\textbf{stable}&
&
\tabularnewline
\hline
$n=5$&
$m=1$&
$m=2$&
$m=3$&
$m=4$&
\tabularnewline
&
unstable&
unstable&
\textbf{stable}&
\textbf{stable}&
\tabularnewline
\hline
$n=6$&
$m=1$&
$m=2$&
$m=3$&
$m=4$&
$m=5$\tabularnewline
&
unstable&
unstable&
unstable&
\textbf{stable}&
\textbf{stable}\tabularnewline
\hline
\end{tabular}

\caption{Table showing the available charges and the stability status of vortex
solutions for different lattice orders. The stable status means that
it is possible to stabilize the vortex under consideration by a suitable
election of the Bessel lattice parameters.\label{cap:Table}}
\end{table}

\begin{figure}

\includegraphics[%
  scale=0.50]{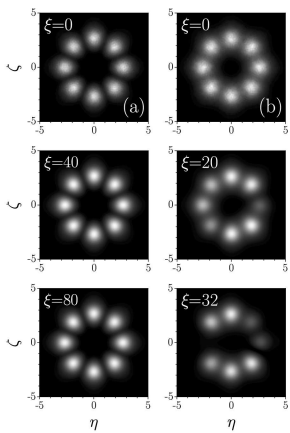}

\caption{Propagation dynamics of vortex solitons with $m=3$ (a) and $m=1$
(b) supported by fourth-order Bessel lattice in the presence of white
input noise with variance $\sigma_{\mathrm{noise}}^{2}=0.01$. Bright regions correspond to high light
intensities, dark regions correspond to low intensities. Both solitons correspond to $b=1.6$ and
$p=28$.\label{cap:propagation}}
\end{figure}

Similar comprehensive linear stability analysis conducted for all lattices
with orders $n$ up to 20 with various depths $0<p<100$
enabled us to derive the important \char`\"{}stability
rule\char`\"{} for vortex solitons, which {\it may be\/} stable
(i.e., a necessary but not sufficient condition) only if the
vortex topological charge satisfies the condition
\begin{equation} \frac{n}{2}<m\le
n-1\label{eq:stability_rule}\end{equation} with the exception for
$n=2$, when the only existing symmetric vortex with charge $m=1$ may be
stable. Similarly to the cases of third-, fourth-, and fifth-order Bessel lattices
(Figs. \ref{cap:diagrams_vortex_charge_1}(d)-(f)),
vortices whose charges satisfy the condition
\ref{eq:stability_rule} become stable in the entire domain of
their existence for deep enough lattices.

Notice that the phase variation $\delta\phi$ between neighboring
lattice maxima for vortices with charges \ref{eq:stability_rule}
verifies $\pi/2<\delta\phi<\pi$. On intuitive grounds, this
difference is consistent with a repulsive interaction between
bright spots forming the vortex (Fig.
\ref{cap:Amplitude_and_phase_vortices}), that are compensated by
the lattice and lead to stable vortex propagation, similarly to
the case of multipole beams \cite{kartashov_et_al}.

Direct numerical integration of Eq.(\ref{eq:evolution_equation})
with input conditions
$q(\eta,\zeta,\xi=0)=w(\eta,\zeta)[1+\rho(\eta,\zeta)]$, where
$w(\eta,\zeta)$ is the stationary solution and $\rho(\eta,\zeta)$
is the white noise with variance
$\sigma_{\mathrm{noise}}^{2}=0.01$, fully confirmed results of
linear stability analysis. In the presence of noisy perturbations,
vortices whose charges satisfy the \char`\"{}stability
rule\char`\"{} propagate undistorted over hundreds of diffraction
lengths, while unstable representatives of vortex soliton families
are rapidly destroyed upon propagation
(Fig.\ref{cap:propagation}). The decay of unstable vortex is accompanied
by progressive increase of intensity oscillations in neighboring
bright spots, until only several spots remain in the output pattern.

In conclusion, we set a general connection between the discrete
rotational symmetry of azymuthally modulated lattices and the
topological winding number of the allowed symmetric vortex
solitons. We derived a \char`\"{}charge rule\char`\"{} and
\char`\"{}stability rule\char`\"{}; the ``charge rule'' is
intended to be general for lattices with a given {\it global\/}
symmetry order and does not depend on the way in which lattice is
created (optical induction in anisotropic nonlinear material or in
Bose-Einstein condensates, direct fabrication in the case of
photonic crystal, or other methods), while the details of
stability of the allowed vortex families are expected to depend on
the {\it local\/} properties of the particular lattice considered.
The rules predict, for example, that in contrast to square
lattices, suitable azimuthally modulated lattices support stable
symmetric vortex solitons with higher-order topological charges.
Because of the general nature of the geometrical method used in
our derivation, we anticipate that a similar charge rule should
hold for scalar soliton systems with comparable azymuthal symmetry
in a variety of physical settings.


This work was partially
supported by contracts BFM2002-2861 and TIC2002-04527-C02-02 from
the Government of Spain, and by Generalitat Valenciana (grants
GV04B-390 and Grupos03/227).

\end{document}